\documentclass[aps,pra,preprint,groupedaddress,showpacs,%
tightenlines,
nofootinbib,floatfix]{revtex4}

\usepackage{amsmath,amssymb,amsfonts,bm}
\usepackage{graphicx}
\usepackage{dcolumn}
\usepackage{mathrsfs} 

\renewcommand{\vec}[1]{\bm{\mathrm{#1}}} %

\begin{document}

\title{Differential cross sections for muonic atom scattering from
  hydrogenic molecules}

\author{Andrzej Adamczak}
\email{andrzej.adamczak@ifj.edu.pl}
\affiliation{Institute of Nuclear Physics, Polish Academy of Sciences,
31-342~Krak\'ow, Poland,\\
Rzesz\'ow Technical University, 35-959~Rzesz\'ow, Poland}

\date{\today}

\begin{abstract}
  The differential cross sections for low-energy muonic hydrogen atom
  scattering from hydrogenic molecules are directly expressed by the
  corresponding amplitudes for muonic atom scattering from
  hydrogen-isotope nuclei. The energy and angular dependence of
  these three-body amplitudes is thus taken naturally into account in
  scattering from molecules, without involving any pseudopotentials.
  Effects of the internal motion of nuclei inside the target molecules
  are included for every initial rotational-vibrational state. These
  effects are very significant as the considered three-body amplitudes
  often vary strongly within the energy interval $\lesssim{}0.1$~eV. The
  differential cross sections, calculated using the presented method,
  have been successfully used for planning and interpreting many
  experiments in low-energy muon physics. Studies of $\mu^{-}$ nuclear
  capture in $p\mu$ and the measurement of the Lamb shift in $p\mu$ atoms
  created in H$_2$ gaseous targets are recent examples.
\end{abstract}


\maketitle

\section{Introduction}
\label{sec:scat_mol_intro}

A~calculation of the differential cross sections for low-energy
scattering of muonic hydrogen atoms from hydrogen-isotope (hydrogenic)
molecules is the main subject of this paper. The cross sections are
expressed in terms of the corresponding amplitudes for muonic atom
scattering from hydrogen-isotope nuclei. Thus, a~dependence of these
three-body scattering amplitudes on the collision energy, scattering
angle, and spin is directly included. For numerical calculations, the
three-body amplitudes computed using the adiabatic
method~\cite{vini82,mele83,buba87,brac89a,brac89,brac90,adam92,chic92},
are employed.

Many experiments in low-energy muon physics are performed using
molecular hydrogen-isotope targets (see,
e.g.,~Refs.~\cite{breu89,pono90,kamm00,kamm01,kott01,pohl01a,pohl05}).
For planning and interpreting such experiments, the differential cross
sections for the following processes are often required:
\begin{subequations}
  \begin{alignat}{2}
    \label{eq:mol_scatt_el}
    &\text{elastic scattering:} \quad 
    & a\mu(F)+BC \to a\mu(F) + BC & , \\
    \label{eq:mol_scatt_ex}
    &\text{isotopic exchange:}          
    & a\mu + BC \to b\mu + AC & , \\
    \label{eq:mol_scatt_sf}
    &\text{and spin-flip:}            
    & a\mu(F)+ AB \to a\mu(F') + AB & .
  \end{alignat}
  \label{eq:mol_scatt}
\end{subequations}
A~monic hydrogen-isotope atom in the $1S$~state is denoted here by
$a\mu$ or~$b\mu$; $F$ and~$F'$ stand for the initial and final total
spin of the muonic atom. The molecules $BC$, $AC$, and $AB$ denote the
hydrogenic molecules H$_2$, D$_2$, T$_2$, HD, HT, or~DT. The
processes~(\ref{eq:mol_scatt}) can take place with simultaneous
rotational and vibrational transitions in a~target molecule. Thus, the
name ``elastic'' assigned here to the scattering~(\ref{eq:mol_scatt_el})
refers solely to the state of the muonic atom. The cross sections for
the processes~(\ref{eq:mol_scatt}) are henceforth called the
``molecular'' cross sections.

For many years, only the cross sections for muonic hydrogen atom
scattering from hydrogen-isotope nuclei (``nuclear'' cross sections)
were available.
\begin{subequations}
  \begin{alignat}{2}
    \label{eq:nuc_scatt_el}
    &\text{elastic scattering:}\qquad 
    & a\mu(F)+b \to a\mu(F)+b & , \\
    \label{eq:nuc_scatt_ex}
    &\text{isotopic exchange:}          
    & a\mu+b \to b\mu+a & , \\
    \label{eq:nuc_scatt_sf}
    &\text{and spin-flip:}            
    & a\mu(F)+a \to a\mu(F')+a & .
  \end{alignat}
  \label{eq:nuc_scatt}
\end{subequations}
The application of the nuclear cross sections to a~description of experiments
performed in molecular targets gives very unsatisfactory results.
A~characteristic kinetic energy of muonic atoms in typical gaseous
targets is lower than a~few~eV~\cite{abbo97}. Therefore, it is necessary
to take into account effects of molecular binding and electron
screening.

Since a~muonic hydrogen atom is a~small neutral system, the methods
developed for the description of neutron scattering in matter can be
adapted, to a~certain extent, for the muonic atom case. Molecular
effects in low-energy neutron scattering from nuclei bound in chemical
compounds are estimated using the Fermi
pseudopotential~\cite{ferm36,brei47,lipp50}. Such a~pseudopotential is
proportional to the constant scattering length. The Fermi method was
used for decades for the calculation of low-energy neutron cross sections
(see e.g., Ref.~\cite{love84} and references therein). In particular,
a~quantum-mechanical treatment of slow neutron scattering from molecular
hydrogen and deuterium was presented by~Young and~Koppel~\cite{youn64}.

A~method of calculating binding effects in the molecular
processes~(\ref{eq:mol_scatt}), based on the Fermi approach, was derived
in~Refs.~\cite{adam89,adam93}. In particular, specific spin-dependent
pseudopotentials were introduced for a~description of a~muonic atom
interaction with a~single nucleus. However, this method has a~limited
applicability since the nuclear processes~(\ref{eq:nuc_scatt}) involve
several partial scattering waves~\cite{brac89,chic92} even at low
($\sim{}1$~eV) energies, in contrast to low-energy neutron scattering.
Moreover, muonic atom scattering often changes strongly (e.g., $p\mu+p$
and~$t\mu+t$) with energy in the intervals comparable with the
rotational thresholds of hydrogenic molecules. A~solution to this
problem is to base a~calculation of the molecular cross sections on the
full nuclear scattering amplitudes, which include all the angular and
energy dependence. The effective radius of interaction between
a~muonic atom and a~nucleus is much smaller than the internuclear
distance in a~hydrogen molecule~\cite{vini82}. Therefore, the amplitude
for scattering from two bound nuclei can be well approximated by a~sum
of the two corresponding amplitudes for scattering from isolated nuclei.
In such an approach, it is necessary to take into account the internal
motion of the nuclei inside a~target molecule. This motion can be
neglected for a~molecule consisting of heavy nuclei. However, we are
dealing with the lightest molecules and, therefore, the kinetic energy of
nuclear motion due to zero-point vibration is on the order of~0.1~eV.

In Sec.~\ref{sec:impulse_approx}, the amplitudes for the molecular
processes~(\ref{eq:mol_scatt}) are expressed in terms of the amplitudes
of the three-body reactions~(\ref{eq:nuc_scatt}). The derived formulas
depend on the momenta of internal motion of the nuclei in a~target
molecule. The differential cross sections for scattering from molecules
are obtained in~Sec.~\ref{sec:difmol}, using a~harmonic model of
molecular vibrations. Also, electron-screening corrections to the cross
sections are given in this section. Some typical examples of the
computed differential cross sections are shown
in~Sec.~\ref{sec:example_mol}.

\section{Amplitudes for scattering from molecules}
\label{sec:impulse_approx}

Let us consider $a\mu$ scattering from a~molecule $BC$ consisting of
hydrogen-isotope nuclei $b$ and~$c$ and two electrons. First, we assume
that the nuclei $b$ and~$c$ are different from the nucleus~$a$, so that
the scattering is spin-independent~\cite{vini82,mele83}. Also, electron
screening effects are neglected in this section. The scattering lengths
of the processes $a\mu+b$~($c$)~\cite{brac89,chic92} are much smaller
than the molecular diameter~$R_0\approx{}300~a_\mu$ ($a_\mu~$~denotes
the Bohr radius of the $a\mu$ atom). The interaction of a~muonic
hydrogen atom with nucleus~$b$ (or~$c$) is important at distances
$\ll{}R_0$~\cite{vini82}. Hence, it is assumed that $a\mu$ interacts
with a~single nucleus during the collision with the molecule. We also assume
that the molecular bond is unperturbed at the moment of collision.
Therefore, $a\mu$ collision with $b(c)$ is treated here as if this
nucleus were free, except for its momentum distribution due to the
molecular binding~\cite{mott65}. This means that the amplitude for
$a\mu$ scattering from a~bound nucleus is the same as that for an
identical free nucleus, provided the momentum of the relative motion is
not changed.

At large distances between $a\mu$ and~$BC$, the initial~$\psi_0$ and
final~$\psi_n$ coordinate wave functions of the system are as follows:
\begin{equation}
  \label{eq:wave_am-BC}
  \begin{split}
    \psi_0(\vec{r},\vec{r}_{\mu},\vec{R}) &=
    \phi_i(\vec{r}_{\mu})\, \varPhi_0(\vec{R}) 
    \exp(i\vec{k}_0\cdot\vec{r})\,, \\
    \psi_n(\vec{r},\vec{r}_{\mu},\vec{R}) &=
    \phi_f(\vec{r}_{\mu})\, \varPhi_n(\vec{R}) 
    \exp(i\vec{k}_n\cdot\vec{r})\,, 
  \end{split}
\end{equation}
where $\vec{k}_0$ and~$\vec{k}_n$ are the initial and final momenta
of~$a\mu$; $\varPhi_0$~and~$\varPhi_n$ are the wave functions of the
initial and final rotational-vibrational states of the molecule~$BC$\@.
The corresponding wave functions of the $1S$~muonic atom are denoted by
$\phi_i$ and~$\phi_f$, where the indices $i$ and~$f$ refer to the
processes~(\ref{eq:nuc_scatt}) with the nuclear scattering
amplitudes~$f_{if}$~\cite{buba87,brac89a,brac89,brac90,adam92,chic92}.
\begin{figure}[htb]
  \begin{center}
    \includegraphics[width=5.5cm]{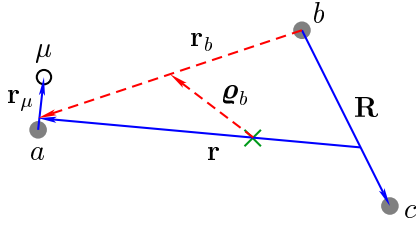}
    \caption{(Color online) Relative coordinates used for the
      description of muonic atom~$a\mu$ scattering from a~molecule~$BC$\@. 
      The cross denotes a~position of the center of mass of this system.
      \label{fig:sys_mol}}
  \end{center}
\end{figure}
In~Fig.~\ref{fig:sys_mol}, the relative coordinates used for providing
a~description of the system are shown. The vector~$\vec{R}$ connects
nucleus~$b$ with nucleus~$c$; $\vec{r}$ denotes the $a\mu$ position with
respect to the center of mass (c.m.) of~$BC$\@; $\vec{r}_{\mu}$~is the
$a\mu$ internal vector. The vector~$\vec{r}_b$ stands for $a\mu$
position relative to nucleus~$b$.

It is convenient to express the amplitude for $a\mu$ scattering
from~$BC$ in terms of the momenta $\vec{k}_0$ and~$\vec{k}_n$ in the
center-of-mass system (c.m.s.) for $a\mu+BC$. On the other hand, the
calculated amplitudes for $a\mu$ scattering from nuclei are functions of
the initial $\vec{p}_b$ and final $\vec{p}_b'$ momenta in the c.m.s.\ of
$a\mu+b$. Therefore, further evaluation of the molecular amplitudes
involves the investigation of a~transition between the ``nuclear'' and
``molecular'' momenta.  First, we assume that both the nuclear and
molecular scattering can be described in the Born approximation. The
amplitude for $a\mu$~scattering from~$b$ bound in~$BC$ is thus given by
the following formula (in muonic atomic units $e=\hbar=\mu_{a\mu}=1$):
\begin{equation}
  \label{eq:amp_imp1}
  \begin{split}
    \mathcal{F}_{0n}^{(b)} (\vec{k}_0,\vec{k}_n) = 
    &-\frac{\mathcal M}{2\pi}\, 
    \int d^3r\, d^3r_{\mu}\, d^3R \, 
    \exp\left[i(\vec{k}_0 -\vec{k}_n)\cdot \vec{r}\,\right]\\
    &\times \varPhi_n^*(\vec{R}) \, \phi_f^*(\vec{r}_{\mu})\,
    V^{(b)}(\vec{r}_b,\vec{r}_{\mu})\, 
    \phi_i(\vec{r}_{\mu})\,\varPhi_0(\vec{R}) \,, 
  \end{split}
\end{equation}
where $\mathcal{M}$~is the reduced mass of the $a\mu+BC$ system (the
masses of the electrons are neglected) and $\mu_{a\mu}$ is the reduced mass
of the $a\mu$~atom
\begin{equation*}
  \begin{split}
    \mathcal M^{-1} & =  M_{a\mu}^{-1} + M_\text{mol}^{-1} \,, \qquad
    \mu_{a\mu}^{-1}   = M_a^{-1}+ m_{\mu}^{-1} , \\
    M_{a\mu} & = M_a + m_{\mu}\,, \qquad\quad M_\text{mol} = M_b + M_c \,.
  \end{split}
\end{equation*}
The potential of $a\mu$~interaction with a~free nucleus~$b$ is denoted
by~$V^{(b)}$. Using the relation
\begin{equation}
  \label{eq:beta}
  \vec{r}_b = \vec{r}+\beta_b\vec{R} \,, \qquad
  \beta_b = M_c/(M_b+M_c) \,,
\end{equation}
in~Eq.~(\ref{eq:amp_imp1}) leads to the following factorization:
\begin{equation}
  \label{eq:amp_imp2}
  \begin{split}
    \mathcal{F}_{0n}^{(b)} = 
    &-\frac{\mathcal M}{2\pi}\, \int d^3r_b\,
    \exp\left[i(\vec{k}_0 -\vec{k}_n)\cdot\vec{r}_b \, \right] 
    V_{if}^{(b)}(\vec{r}_b) \\
    \times &\int d^3R \, 
    \exp\left[i\beta_b (\vec{k}_0 -\vec{k}_n)\cdot\vec{R}\, \right] 
    \varPhi_n^*(\vec{R}) \varPhi_0(\vec{R}) \,,
  \end{split}
\end{equation}
in which $V_{if}^{(b)}$ denotes the ``nuclear'' matrix element   
\begin{equation}
  \label{eq:V_if-b}
  V_{if}^{(b)}(\vec{r}_b) \equiv \int d^3r_{\mu}\,
  \phi_f^*(\vec{r}_{\mu})\, V^{(b)}(\vec{r}_b,\vec{r}_{\mu})
  \phi_i(\vec{r}_{\mu}) \,.
\end{equation}
The first integral in~Eq.~(\ref{eq:amp_imp2}) is the Born amplitude for
$a\mu$ scattering from a~free nucleus~$b$ times the
factor~$\mathcal{M}/\mu_b$, where~$\mu_b$ stands for the reduced mass of
the system $a\mu+b$:
\begin{equation*}
  \mu_b^{-1} = (m_{\mu}+M_a)^{-1} + M_b^{-1} .
\end{equation*}
The second integral in~Eq.~(\ref{eq:amp_imp2}) is a~form factor
describing the binding of~$b$ in~$BC$\@.

In order to investigate a~dependence of the molecular
amplitude~$\mathcal{F}_{0n}^{(b)}$ on the internal motion of~$b$ inside
the molecule, a~momentum representation of the wave function~$\varPhi_n$
is introduced
\begin{equation}
  \label{eq:wav_mol_Fourier}
  \begin{split}
    &\varPhi_n(\vec{R}) \equiv \frac{1}{(2\pi)^{3/2}}\int d^3\kappa_n \, 
    \exp(i\beta_b\vec{\kappa}_n \cdot \vec{R}) \, 
    g_n^{(b)}(\vec{\kappa}_n) , \\
    &g_n^{(b)}(\vec{\kappa}_n) \equiv \frac{\beta_b^3}{(2\pi)^{3/2}} 
   \int d^3R \, \exp(-i\beta_b\vec{\kappa}_n 
    \cdot \vec{R}) \, \varPhi_n(\vec{R}) .
  \end{split}
\end{equation}
The vector~$\vec{\kappa}_n$ is the momentum of the internal nuclear
motion in the final rotational-vibrational state~$n$. The analogous
equations can be written down for the initial molecular
state~$\varPhi_0$ with the internal nuclear momentum~$\vec{\kappa}_0$.
Upon, substituting~Eq.~(\ref{eq:wav_mol_Fourier})
into~Eq.~(\ref{eq:amp_imp1}) one obtains:
\begin{equation}
  \label{eq:amp_imp3}
  \begin{split}
    \mathcal{F}_{0n}^{(b)} = 
    -\frac{\mathcal M}{(2\pi)^4} \int & d^3r\, d^3R \, 
    d^3\kappa_n\, d^3\kappa_0\,
    \exp\left[i(\vec{k}_0 -\vec{k}_n) \cdot \vec{r}\, \right]\, \\
    &\times
    \exp\left[i\beta_b (\vec{\kappa}_0 -\vec{\kappa}_n) \cdot 
    \vec{R}\, \right] 
    V_{if}^{(b)}(\vec{r}_b)\, \\
    &\times
    g_n^{(b)*}(\vec{\kappa}_n)\, g_0^{(b)}(\vec{\kappa}_0)\,.
  \end{split}
\end{equation}
Then, using new variables $\vec{r}_b$ and~$\vec{\varrho}_b$
\begin{equation}
  \label{eq:vec_r_rho_b}
  \begin{split}
    &\vec{r} = \frac{\mu_b}{\mathcal{M}}\, \vec{r}_b 
    + \beta_b \frac{M_{abc}}{M_c}\, \vec{\varrho}_b \,,
    \quad 
    \vec{R} = \frac{M_a}{M_{ab}}\, \vec{r}_b 
    -\frac{M_{abc}}{M_c}\, \vec{\varrho}_b \,, \\
    &M_{abc} = M_a+M_b+M_c \,, \quad M_{ab} = M_a+M_b \,,
  \end{split}
\end{equation}
in Eq.~(\ref{eq:amp_imp3}) and performing integration over the vector
$\vec{x}=(\beta_b{}M_{abc}/M_c)\vec{\varrho}_b$, one has
\begin{widetext}
\begin{equation}
  \label{eq:amp_imp5}
  \begin{split}
    \mathcal{F}_{0n}^{(b)} = 
    -\frac{\mathcal M}{2\pi\beta_b^3} 
     \int d^3\kappa_n\, d^3\kappa_0 \, \int & d^3r_b\, 
     \exp\left[-i\left( \frac{\mu_b}{\mathcal{M}}\,\vec{k}_n
        +\beta_b \frac{\mu_b}{M_b}\,\vec{\kappa}_n \right)
      \cdot \vec{r}_b \right]  \\
    &\times V_{if}^{(b)}(\vec{r}_b)\,
    \exp\left[i\left( \frac{\mu_b}{\mathcal{M}}\,\vec{k}_0
        +\beta_b \frac{\mu_b}{M_b}\,\vec{\kappa}_0 \right)
      \cdot \vec{r}_b \right]   \\
    &\times
    \delta (\vec{k}_0 -\vec{k}_n -\vec{\kappa}_0+\vec{\kappa}_n )\, 
    g_n^{(b)*}(\vec{\kappa}_n)\, g_0^{(b)}(\vec{\kappa}_0) \,.
  \end{split}
\end{equation}
The integral over~$\vec{r}_b$ times $-\mu_b/2\pi$ is the Born
amplitude~$f_{if}^{(b)}$ for $a\mu$ scattering on a~free nucleus~$b$,
expressed by the momenta $\vec{k}_n$, $\vec{\kappa}_n$, $\vec{k}_0$,
and~$\vec{\kappa}_0$. Thus, Eq.~(\ref{eq:amp_imp5}) can be written down
in the following form:
\begin{equation}
  \label{eq:amp_imp6}
  \begin{split}
    \mathcal{F}_{0n}^{(b)} = 
    \frac{1}{\beta_b^3}\,\frac{\mathcal{M}}{\mu_b} &
    \int d^3\kappa_n\, d^3\kappa_0 \, f_{if}^{(b)}
    \left( \frac{\mu_b}{\mathcal{M}}\,\vec{k}_0
      +\beta_b \frac{\mu_b}{M_b}\,\vec{\kappa}_0 \,,
      \frac{\mu_b}{\mathcal{M}}\,\vec{k}_n
        +\beta_b \frac{\mu_b}{M_b}\,\vec{\kappa}_n \right) \\
    &\times
    \delta (\vec{k}_0 -\vec{k}_n -\vec{\kappa}_0+\vec{\kappa}_n )\, 
    g_n^{(b)*}(\vec{\kappa}_n)\, g_0^{(b)}(\vec{\kappa}_0) \,,
  \end{split}
\end{equation}
where subscripts $i$ and~$f$ label the kind of nuclear
process~(\ref{eq:nuc_scatt}). Now, we make the basic assumption that
Eq.~(\ref{eq:amp_imp6}) is fulfilled by the \textit{exact} nuclear
amplitudes~$f_{if}^{(b)}$. The integration over~$\vec{\kappa}_n$ is
performed readily using the conservation of the total momentum, which
gives:
\begin{align}
  \label{eq:amp_imp7}
  \mathcal{F}_{0n}^{(b)} (\vec{k}_0,\vec{q}) &= 
  \frac{1}{\beta_b^3}\,\frac{\mathcal{M}}{\mu_b} 
  \int d^3\kappa_0\, 
  f_{if}^{(b)}(\vec{p}_b\,, \vec{p}_b+\vec{q})\,
  g_n^{(b)*}(\vec{\kappa}_0+\vec{q})\, g_0^{(b)}(\vec{\kappa}_0) \,, \\
  \label{eq:p_b}
  \vec{p}_b &\equiv \frac{\mu_b}{\mathcal{M}}\,\vec{k}_0
  +\beta_b \frac{\mu_b}{M_b}\,\vec{\kappa}_0 \,, \qquad
  \vec{p}_b' \equiv \vec{p}_b+\vec{q} \,. 
\end{align}
\end{widetext}
The vector~$\vec{q}$ denotes the momentum transfer
\begin{equation}
  \label{eq:q_transfer}
  \vec{q} = \vec{k}_n - \vec{k}_0 = \vec{p}_b' - \vec{p}_b \,.
\end{equation}
Thus, the molecular amplitude~$\mathcal{F}_{0n}^{(b)}$ for scattering
with a~fixed momentum transfer~$\vec{q}$ is determined by the free
nuclear amplitude~$f_{if}^{(b)}$ with the same momentum transfer.
However, the initial momentum~$\vec{p}_b$ in the $a\mu+b$ c.m.s.\ is
different from the initial momentum~$\vec{k}_0$ in the molecular c.m.s..
According to~Eq~(\ref{eq:p_b}), the vector~$\vec{p}_b$ depends also on
the internal motion of~$b$. This gives the following $a\mu$ kinetic
energy in the $a\mu+b$ c.m.s.:
\begin{equation}
  \label{eq:e_b}
  \varepsilon_b = \frac{\mu_b}{\mathcal{M}}\, \varepsilon
  + \beta_b^2 \frac{\mu_b}{\mu_{bc}}\, \varepsilon_{bc} 
  + 2\beta_b \frac{\mu_b}{\sqrt{\mu_{bc}\,\mathcal{M}}}\,
  \sqrt{\varepsilon\, \varepsilon_{bc}}\,\cos\theta \,, 
\end{equation}
where $\theta$ is the angle between the vectors $\vec{k}_0$
and~$\vec{\kappa}_0$. The muonic atom kinetic energies in the
nuclear~($\varepsilon_b$) and molecular~($\varepsilon$) c.m.s.\ are
\begin{equation}
  \label{eq:def_e_b}
  \varepsilon_b = \frac{p_b^2}{2\mu_b} \,, \qquad 
  \varepsilon = \frac{k_0^2}{2\mathcal{M}} \,.
\end{equation}
The internal kinetic energy~$\varepsilon_{bc}$ of the molecule~$BC$\@ is
\begin{equation}
  \label{eq:def_e_bc}
   \varepsilon_{bc} = \frac{\kappa_0^2}{2\mu_{bc}} \,, \qquad
   \mu_{bc}^{-1} = M_b^{-1} + M_c^{-1}.
\end{equation}
At $\varepsilon\to{}0$, the collision energy~$\varepsilon_b$ in the
$a\mu$+b system is determined solely by~$\varepsilon_{bc}$. This energy
never vanishes because of the zero-point vibration of the molecule. In
particular, for the lightest H$_2$ molecule, the second term
of~Eq.~(\ref{eq:e_b}) is on the order of~$0.01$~eV. This energy is
inaccessible, but it affects the molecular scattering
amplitude~(\ref{eq:amp_imp7}). For a~fixed~$\varepsilon$, the spectrum
of~$\varepsilon_b$ is quite wide. Its width is determined by the
term~$\sqrt{\varepsilon\,\varepsilon_{bc}}$ which depends on the
hydrogenic-molecule vibrational quantum ($\approx{}0.3$--0.5~eV).
Therefore, at a~given~$\varepsilon$, the molecular
amplitude~(\ref{eq:amp_imp7}) contains contributions from the nuclear
amplitude~$f_{if}$ taken at different energies. This effect should be
taken into account when~$f_{if}$ changes significantly within the
spectrum~(\ref{eq:e_b}) of~$\varepsilon_b$, which often occurs in muonic
atom scattering.

\begin{figure}[htb]
  \begin{center}
    \includegraphics[width=4.8cm]{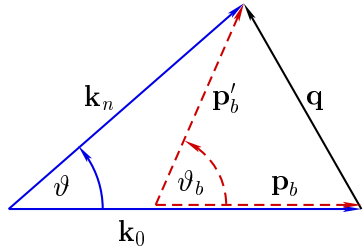}
    \caption{(Color online) Relations between the initial and final 
      momenta and between the scattering angles $\vartheta_b$ 
      and~$\vartheta$ in the nuclear and molecular c.m.s.\ 
      (for $k_0\gg\kappa_0$).
    \label{fig:mol_q}}
  \end{center}
\end{figure}
If $k_0\gg\kappa_0$ and $k_n\gg\kappa_n$, the dependence of~$\vec{p}_b$
on~$\vec{\kappa}_0$ in~Eq.~(\ref{eq:p_b}) can be neglected. As a~result,
we get:
\begin{equation}
  \label{eq:p_b_asymp}
  \vec{p}_b \approx \frac{\mu_b}{\mathcal{M}}\,\vec{k}_0 
\end{equation}
and Eq.~(\ref{eq:amp_imp7}) is then factorized as follows: 
\begin{equation}
  \label{eq:amp_imp8}
  \mathcal{F}_{0n}^{(b)} = 
  \frac{1}{\beta_b^3}\,\frac{\mathcal{M}}{\mu_b} \,
  f_{if}^{(b)}(\vec{p}_b\,, \vec{p}_b+\vec{q})
  \int d^3\kappa_0\, 
  g_n^{(b)*}(\vec{\kappa}_0+\vec{q})\, g_0^{(b)}(\vec{\kappa}_0) \,. 
\end{equation}
This formula can also be used when $f_{if}^{(b)}$ weakly depends on
the variation of~$p_b$ due to the characteristic spectrum
of~$\vec{\kappa}_0$. After the substitution of the Fourier
transforms~(\ref{eq:wav_mol_Fourier}) into Eq.~(\ref{eq:amp_imp8}) and
integration over~$\vec{\kappa}_0$, we obtain
\begin{equation}
  \label{eq:amp_imp10} 
  \begin{split}
    \mathcal{F}_{0n}^{(b)} = &
    \frac{\mathcal{M}}{\mu_b} \,
    f_{if}^{(b)}\left( \frac{\mu_b}{\mathcal{M}}\,\vec{k}_0
    \,, \frac{\mu_b}{\mathcal{M}}\,\vec{k}_0 +\vec{q}\right)\\
    &\times \int d^3R\; \varPhi_n^*(\vec{R}) 
    \exp(i\beta_b\vec{q}\cdot\vec{R})\, \varPhi_0(\vec{R}) \,.
  \end{split}
\end{equation}
The molecular scattering amplitudes~(\ref{eq:amp_imp7})
and~(\ref{eq:amp_imp10}) naturally take into account the dependence of
the nuclear amplitude~$f_{if}$ on the $a\mu$ energy and scattering
angle~$\vartheta$. At low collision energies, the amplitudes~$f_{if}$
are well approximated by the corresponding constant scattering
lengths~$\lambda_{if}^0$. As a~result, Eq.~(\ref{eq:amp_imp10}) is
simplified:
\begin{equation}
  \label{eq:amp_imp_len}
  \begin{split}  
    \mathcal{F}_{0n}^{(b)}(\vec{q}) &\approx
    -\frac{\mathcal{M}}{\mu_b} \, \lambda_{if}^{(b)0}
    \int d^3R\; \varPhi_n^*(\vec{R}) 
    \exp(i\beta_b \vec{q}\cdot\vec{R})\, \varPhi_0(\vec{R}), \\ 
    \mathcal{F}_{0n}^{(b)}(\vec{q}) &\xrightarrow[~~q\to{}0~~] \,
    -\frac{\mathcal{M}}{\mu_b} \, \lambda_{if}^{(b)0}.
  \end{split}
\end{equation}
This equation can be formally obtained in the first Born approximation,
using the pseudopotential
\begin{equation}
  \label{eq:pseud_s}
  V_{if}^{(b)}(\vec{r}_b) = \frac{2\pi}{\mu_b} \, 
  \lambda_{if}^{(b)0}\,
  \delta(\vec{r}_b) = \frac{2\pi}{\mu_b} \, 
  \lambda_{if}^{(b)0}\,
  \delta(\vec{r}+\beta_b \vec{R})    
\end{equation}
in Eq.~(\ref{eq:amp_imp1}). Such a~potential for a~constant scattering
length was first introduced by Fermi~\cite{ferm36} and more rigorously
derived by Breit~\cite{brei47} and then by Lippmann and
Schwinger~\cite{lipp50}.

In the case of a~general spherical potential with a~finite range, it is
possible to generalize the Fermi pseudopotential by the introduction of
partial pseudopotentials corresponding to subsequent scattering
waves~\cite{huan57,stoc05}. However, the calculations of the cross
sections for muonic atom scattering on molecules, presented in this
paper, are directly based on a~knowledge of the amplitudes for nuclear
scattering.  Therefore, a~formulation of a~generalized pseudopotential,
in this case, is superfluous. Let us only note that the correct nuclear
partial amplitudes can be formally obtained by the substitution of the
following pseudopotentials:
\begin{equation}
  \label{eq:pseud_partial}
  V_{if}^{(b)J}(\vec{r}_b) = \frac{2\pi}{\mu_b} \, (2J+1)\, 
  \lambda_{if}^{(b)J}(p_b) \, \delta(\vec{r}_b)\, 
  \text{P}_J(\cos\vartheta_b) 
\end{equation}
into Eq.~(\ref{eq:amp_imp1}). The energy-dependent nuclear scattering
length $\lambda_{if}^{(b)J}$ is defined as follows:
\begin{equation}
  \label{eq:part_scatt_length}
  f_{if}^{(b)J} = - (2J+1)\, \lambda_{if}^{(b)J} \,
  \text{P}_J(\cos\vartheta_b)  \,,
\end{equation}
where $f_{if}^{(b)J}$ are the partial nuclear amplitudes for $a\mu+b$
scattering~\cite{buba87,brac89a,brac89,brac90,adam92,chic92} and $J$ is
the angular momentum of the $a\mu+b$ system.  The angle between the
vectors $\vec{p}_b$ and~$\vec{p}_b'$ is denoted here by~$\vartheta_b$.
The function P$_J$ is the $J$th~Legendre polynomial.

The problem of the angular and energy dependence of the nuclear
scattering $t\mu+d$ in the $t\mu+\text{D}_2$ process was alternatively
solved in~\cite{bouk99} by the introduction of the effective
polarization potential ($\sim{}r^{-4}$). However, the magnitude of such
a~potential for the $t\mu$-$d$~interaction was determined separately for
every given collision energy. Thus, such an approach is more complicated
than the direct use of the computed nuclear-scattering amplitudes and
neglects a~wide distribution of the deuteron kinetic energies in the
D$_2$~molecule.

When the internal motion of the nuclei inside the target molecule cannot
be neglected, the molecular amplitude is given by
Eq.~(\ref{eq:amp_imp7}).  However, in a~general case, the numerical
evaluation of the integrals over~$\vec{\kappa}_0$ is difficult. The role
of the internal motion is most important if the condition
$q\ll{}\kappa_0$ is fulfilled, which implies that the internal state of
the molecule is not changed ($g_n^{(b)}=g_0^{(b)}$). In this case,
Eq.~(\ref{eq:amp_imp7}) is approximated as follows:
\begin{equation}
  \label{eq:amp_avg1}
  \begin{split}
    \mathcal{F}_{00}^{(b)}(k_0) & \approx \frac{\mathcal{M}}{\mu_b} \,
    \overline{f_{if}^{(b)}}(k_0) \,, \\
    \overline{f_{if}^{(b)}}(k_0) & \equiv 
    \int d^3\kappa_0\, f_{if}^{(b)}(\vec{p}_b,\vec{p}'_b)\, 
    \mathcal{P}_0^{(b)}(\vec{\kappa_0}) \,,
 \end{split}
\end{equation}
with $\mathcal{P}_0^{(b)}(\vec{\kappa_0})$ being a~distribution of the
momentum of nucleus~$b$ in the molecule:
\begin{equation}
  \label{eq:dis_kappa0}
  \mathcal{P}_0^{(b)}(\vec{\kappa_0}) \equiv \beta_b^{-3}\, 
  \lvert g_0^{(b)}(\vec{\kappa_0}) \rvert^2 .
\end{equation}
For elastic scattering at $k_0\to{}0$, when one expects that the
internal-motion effect is the strongest, the molecular scattering
amplitude is
\begin{equation}
  \label{eq:amp_avg_s}
  \mathcal{F}_{00}^{(b)} \approx -\frac{\mathcal{M}}{\mu_b} \;
  \overline{\lambda_{if}^{(b)0}} ,
\end{equation}
where the bar denotes averaging over~$\vec{\kappa}_0$.
Equations~(\ref{eq:amp_avg1}) and~(\ref{eq:amp_avg_s}) suggest
a~reasonable approximation of formula~(\ref{eq:amp_imp7}) for
finite~$k_0$. When the exact nuclear amplitude~$f_{if}^{(b)}$ is
replaced by the averaged function
\begin{equation}
  \label{eq:amp_avg_J}
  \begin{split}
    \overline{f_{if}^{(b)}}(\vec{k}_0,\vec{q}) &\equiv 
    -\sum_J (2J+1)\, \overline{\lambda_{if}^{(b)J}}(k_0)\,
    \text{P}_J(\cos\vartheta_b) \,, \\
    \overline{\lambda_{if}^{(b)J}}(k_0) &\equiv 
    \int d^3\kappa_0\,\,
    \lambda_{if}^{(b)J}(p_b)\, 
    \mathcal{P}_0^{(b)}(\vec{\kappa_0}) \,,
  \end{split}
\end{equation}
Eq.~(\ref{eq:amp_imp7}) is factorized. This leads, finally, to an
equation similar to~Eq.~(\ref{eq:amp_imp10}), with $f_{if}^{(b)}$
replaced by the mean amplitude~$\overline{f_{if}^{(b)}}$. This
approximation gives the limit~(\ref{eq:amp_avg_s}) at $k_0\to{}0$. On
the other hand, this approximation coincides with the asymptotic
amplitude~(\ref{eq:amp_imp10}) at $k_0\gg{}\kappa_0$. The dependence
of~$\vartheta_b$ on~$\vec{\kappa}_0$ is neglected here since the higher
partial waves ($J>0$) in the nuclear scattering are important only at
$k_0\gg{}\kappa_0$.

\section{Molecular differential cross sections}
\label{sec:difmol}

\subsection{Spin-independent scattering}
\label{sec:difmol_asym}

In the presented approach, the total amplitude~$\mathcal{F}_{0n}$ for
$a\mu$ scattering on a~molecule $BC$\@ is equal to the sum of the
amplitudes for scattering on the bound nuclei~$b$ and~$c$
\begin{equation}
  \label{eq:amp_mol1}
  \mathcal{F}_{0n}(\vec{k}_0,\vec{q}) =
  \mathcal{F}_{0n}^{(b)}(\vec{k}_0,\vec{q}) +
  \mathcal{F}_{0n}^{(c)}(\vec{k}_0,\vec{q}) \,, 
\end{equation}%
where $\mathcal{F}_{0n}^{(b)}$ is given by~Eq.~(\ref{eq:amp_imp10}), and
the derivation of~$\mathcal{F}_{0n}^{(c)}$ is analogous. Let us first
consider the spin-independent case $a\neq{}b$,~$c$. Assuming that
vibrations of the molecule are harmonic and that there is no coupling
between the vibrational and rotational degrees of freedom, the molecular
wave function~$\varPhi_n(\vec{R})$ takes the form:
\begin{equation}
  \label{eq:wave_fun_mol}
  \varPhi_n(\vec{R}) = \frac{u_\nu(R)}{R}\; 
  \text{Y}_{KM_K}(\hat{R}) \,,  \qquad  
  \hat{R}\equiv \frac{\vec{R}}{R} \,,
\end{equation}
where quantum numbers $K$,~$M_K$ label the rotational state of~$BC$\@.
The radial wave function~$u_{\nu}$ corresponding to vibrational quantum
number~$\nu$ is
\begin{equation}
  \label{eq:radial_mol}
  \begin{split}
    u_\nu(R) &= \mathcal{N}_{\nu} \,
    \text{H}_{\nu}\bigl[\alpha(R-R_0)\bigr] 
    \exp\left[-\tfrac{1}{2} \alpha^2 (R-R_0)^2\right] \,, \\
    \mathcal{N}_{\nu} &=
    \sqrt{\frac{\alpha}{\sqrt{\pi}\, 2^{\nu} \nu!}}\,,\qquad   
    \alpha = \sqrt{\mu_{bc}\,\omega_0},
  \end{split}
\end{equation}
where $\text{H}_{\nu}$ denotes the $\nu$th Hermite polynomial. The
rotational~$E_K$ and vibrational~$E_{\nu}$ energy levels are given as
\begin{equation}
  \label{eq:energy_mol}
  E_K = B_\text{rot}\,K (K+1) \,, \qquad 
  E_{\nu} = \bigl(\nu+\tfrac{1}{2}\bigr)\,\omega_0 \,.
\end{equation}
At the temperatures usually applicable to experiments, hydrogenic
molecules are initially in the ground vibrational state~$\nu=0$.

Inserting the expansion of the free-wave function (in terms of the
spherical Bessel functions~$\text{j}_l$ and the spherical
harmonics~$\text{Y}_{lm}$) into~Eq.~(\ref{eq:amp_imp10}), one obtains
for the bound nucleus~$b$
\begin{equation}
  \label{eq:amp_mol_expan1}
  \begin{split}
    \mathcal{F}_{0n}^{(b)} = & 4\pi\, \frac{\mathcal{M}}{\mu_b}\,
    f_{if}^{(b)} \sum_{l,m} i^l\, \mathcal{D}_{\nu l}(\beta_b q) \,
    \text{Y}_{lm}^*(\hat{q}) \\
    &\times \int d\varOmega_R \; 
    \text{Y}_{K'M_K'}^*(\hat{R})\, \text{Y}_{lm}(\hat{R})\,  
    \text{Y}_{KM_K}(\hat{R}) \,.
  \end{split}
\end{equation}
The real function $\mathcal{D}_{\nu l}(\beta_b q)$ is a~result of the
integration over~$R$
\begin{equation}
  \label{eq:D_nu_l}
  \begin{split}
    \mathcal{D}_{\nu l}(\beta_b q) &\equiv \int_{0}^{\infty} dR\, 
    u_{\nu}(R)\, \text{j}_l(\beta_b q R)\, u_0(R) \,, \\
    \mathcal{D}_{\nu l}(\beta_b q)  &\xrightarrow[q\to\, 0]{}  
    \begin{cases}
      1&  \text{ if } \nu=0 \text{ and } l=0  \\
      0&  \text{ otherwise.}
    \end{cases}
  \end{split}
\end{equation}
The initial state of the molecule is denoted here by the set of
rotational and vibrational quantum numbers~$0$=$(K,M_K,\nu$=$0)$. The
final state is labeled by~$n$=$(K',M_K',\nu)$. The indices~$i$ and~$f$
in the nuclear amplitude~$f_{if}^{(b)}$ refer to the initial and final
states (spin or isotopic) of the scattered muonic atom.  Integration of
the three spherical harmonics in~Eq.~(\ref{eq:amp_mol_expan1}) over the
solid angle~$\varOmega_R$ leads to the following result:
\begin{equation}
  \label{eq:amp_mol_expan2}
  \begin{split}
    \mathcal{F}_{0n}^{(b)} = & 
    \bigl[ 4\pi (2K'+ 1)(2K+1)\bigr]^{1/2} \, 
    (-1)^{M_K'} \, i^{K-K'} \,\\ 
    & \times \frac{\mathcal{M}}{\mu_b} \,
    f_{if}^{(b)} \sum_{l,m} \, (2l+1)^{1/2}  
    \mathcal{D}_{\nu l}(\beta_b q) \, 
    \text{Y}_{lm}^*(\hat{q}) \\
    &\times 
    \begin{pmatrix}
      K' & l & K \\
      0  & 0 & 0
    \end{pmatrix}
    \begin{pmatrix}
      K' & l & K \\
      -M_K'  & m & M_K
    \end{pmatrix} ,
  \end{split}
\end{equation}
expressed by the Wigner $3j$~symbols.

The molecular differential cross section, averaged over the
projection~$M_K$ of the initial angular momentum and summed over the
projection~$M_K'$ of the final angular momentum, is equal to
\begin{equation}
  \label{eq:x_mol1}
  \frac{d\sigma_{0n}}{d\varOmega} =
  \frac{k_n}{k_0}\, \frac{1}{2K+1}\, \sum_{M_K,M_K'} 
  \left| \mathcal{F}_{0n}^{\text{~mol}}\right|^2 .
\end{equation}
The solid angle~$\varOmega(\vartheta,\varphi)$ is connected with the
direction of the vector~$\vec{k}_n$ with respect to the initial $a\mu$
momentum~$\vec{k}_0$ (see Fig.~\ref{fig:mol_q}). Substitution of
Eq.~(\ref{eq:amp_mol_expan2}) and the analogous formula for the
nucleus~$c$ into~Eqs.~(\ref{eq:amp_mol1}) and~(\ref{eq:x_mol1}) gives
the following cross section:
\begin{equation}
  \label{eq:x_mol_asym}
  \begin{split}
    \frac{d\sigma_{0n}}{d\varOmega} = & \frac{k_n}{k_0}  
    \sum_l \mathcal{W}_{K'lK}\, \Biggl[
    \left(\dfrac{\mathcal{M}}{\mu_b}\right)^2 
   \bigl| f_{if}^{(b)}\bigr|^2\, \mathcal{D}_{\nu l}^2(\beta_b q)\\
   &+(-1)^l\,2 \dfrac{\mathcal{M}^2}{\mu_b\,\mu_c}
   \Re \left(f_{if}^{(b)*}f_{if}^{(c)}\right)
   \mathcal{D}_{\nu l}(\beta_b q)\, \mathcal{D}_{\nu l}(\beta_c q)\\
   &+\left(\dfrac{\mathcal{M}}{\mu_c}\right)^2 
   \bigl| f_{if}^{(c)}\bigr|^2\, \mathcal{D}_{\nu l}^2(\beta_c q) 
   \Biggr] ,
  \end{split}
\end{equation}
where the angular-momentum factor~$\mathcal{W}_{K'lK}$ is defined as
\begin{equation}
  \label{eq:Wig_K'lK}
  \mathcal{W}_{K'lK} \equiv (2K'+1)(2l+1)
      \begin{pmatrix}
      K' & l & K \\
      0  & 0 & 0
    \end{pmatrix}^2 ,  \quad
     \mathcal{W}_{K0K} = 1 \,.
\end{equation}
The reduced mass of the $a\mu+c$ system is
\begin{equation*}
  \mu_c^{-1} = (m_{\mu}+M_a)^{-1} + M_C^{-1} 
\end{equation*}
and
\begin{equation*}
  \beta_c = 1 - \beta_b = M_b/(M_b+M_c) .
\end{equation*}
When $BC$\@ is symmetric ($b=c$, $\mu_b=\mu_c=\mu$,
$\beta_b=\beta_c=\beta=\tfrac{1}{2}$, and
$f_{if}^{(b)}=f_{if}^{(c)}=f_{if}$), Eq.~(\ref{eq:x_mol_asym}) takes the
simpler form:
\begin{equation}
  \label{eq:x_mol_sym1}
  \begin{split}
    \frac{d\sigma_{0n}}{d\varOmega} = 
    2 \left(\dfrac{\mathcal{M}}{\mu}\right)^2  
    \bigl| f_{if}\bigr|^2 \frac{k_n}{k_0}
    \sum_l & \left[(-1)^l+1\right]\, \\
    &\times \mathcal{W}_{K'lK} \, \mathcal{D}_{\nu l}^2(\beta q) \,.
  \end{split}
\end{equation}
The molecular cross sections~(\ref{eq:x_mol_asym})
and~(\ref{eq:x_mol_sym1}) directly include the dependence of the
``bare'' nuclear amplitudes~$f_{if}$ on the collision
energy~$\varepsilon_b$~($\varepsilon_c$) and on the scattering
angle~$\vartheta_b$~($\vartheta_c$). They are derived for high collision
energies~$k_0\gg\kappa_0$. However, they can also be used at lower
energies as a~reasonable approximation if the nuclear
amplitudes~$f_{if}^{(b)}$ ($f_{if}^{(c)}$) are replaced by the
amplitudes~$\overline{f_{if}^{(b)}}$ ($\overline{f_{if}^{(c)}}$)
averaged over~$\mathcal{P}_0^{(b)}$ ($\mathcal{P}_0^{(c)}$). In the case
of the wave function~(\ref{eq:wave_fun_mol}), the momentum
distribution~(\ref{eq:dis_kappa0}) for the ground vibrational
state~$\nu$=$0$ of the molecule~$BC$\@ has the form
\begin{equation}
  \label{eq:dis_kappa0_harmonic}
  \begin{split}
    \mathcal{P}_0^{(b)}(\vec{\kappa}_0) = & \,
    4 \frac{\beta_b^3\, R_0^2}{\alpha\, \sqrt{\pi}}\:
    \text{j}_K^2 (\beta_b\kappa_0 R_0)
    \exp\left(-\dfrac{\beta_b^2\kappa_0^2}{\alpha^2}\right) \\
    &\times \bigl|\text{Y}_{KM_K}(\hat{\kappa}_0)\bigr|^2 ,  
  \end{split}
\end{equation}
in which $\hat{\kappa}_0=\vec{\kappa}_0/\kappa_0$. After averaging
$\mathcal{P}_0^{(b)}$ over orientations of the molecule, one obtains
\begin{equation}
  \label{eq:dis_kappa0_abs}
  \mathcal{P}_0^{(b)}(\kappa_0) = 
   \frac{\beta_b^3\, R_0^2}{\alpha\, \pi^{3/2}}\:
   \text{j}_K^2 (\beta_b\kappa_0 R_0)
   \exp\left(-\dfrac{\beta_b^2\kappa_0^2}{\alpha^2}\right).
\end{equation}
A~distribution~$\mathcal{P}_{0K}$ of the internal kinetic
energy~$\varepsilon_{bc}$ of the target molecule can be derived
similarly. For~$\nu=0$, one has
\begin{equation}
  \label{eq:dis_rovib}
  \begin{split}
    \mathcal{P}_{0K}(\varepsilon_{bc})\, d\varepsilon_{bc} = &
    \frac{2R_0^2\alpha^2}{\pi}\,
    \text{j}_K^2\bigl(R_0\alpha\sqrt{\omega_{bc}}\,\bigr)
    \exp(-\omega_{bc}) \\ 
    & \times \sqrt{\omega_{bc}}\; d\omega_{bc} \,, 
  \end{split}
\end{equation}
where $\omega_{bc}=2\varepsilon_{bc}/\omega_0$.
This distribution is widest for the lightest H$_2$ molecule. According
to~Eq.~(\ref{eq:e_b}), this leads to a~broad distribution of the
collision energy~$\varepsilon_b$ in the nuclear~c.m.s. for a~fixed
collision energy~$\varepsilon$ in the molecular~c.m.s..
\begin{figure}[htb]
  \begin{center}
    \includegraphics[width=7cm]{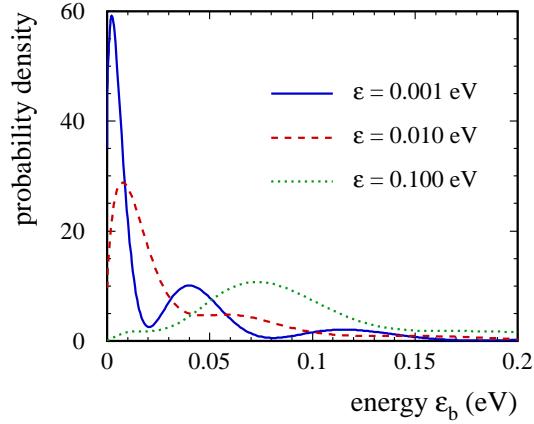}
    \caption{(Color online) Distribution of collision 
      energy~$\varepsilon_b$ in the $p\mu+p$ c.m.s., 
      for a~fixed collision energy~$\varepsilon$ in the 
      $p\mu+$H$_2(K=0)$ c.m.s..
      \label{fig:enucdis}}
  \end{center}
\end{figure}
\begin{figure}[htb]
  \begin{center}
    \includegraphics[width=7cm]{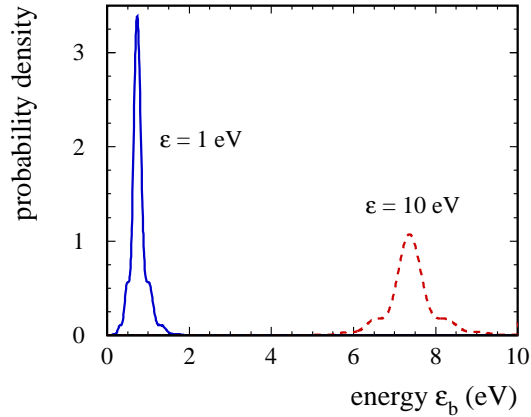}
    \caption{(Color online) The same as in Fig.~\ref{fig:enucdis},
      for $\varepsilon=1$ and~10~eV.
    \label{fig:enucdis2}}
  \end{center}
\end{figure}
In~Figs.~\ref{fig:enucdis} and~\ref{fig:enucdis2}, the calculated
$\varepsilon_b$~spectrum for the ground-state H$_2$ molecule, at several
values of~$\varepsilon$ is presented. At the lowest energies, the shape
of this spectrum is mainly determined by the motion of a~bound proton.
The mean value of~$\varepsilon_b$ equals 0.036~eV for
$\varepsilon=0.001$~eV. For $\varepsilon=0.1$~eV, a~single broad peak
with a~mean value of~0.11~eV is observed in the
$\varepsilon_b$~spectrum. Only at $\varepsilon\gtrsim$~1~eV, does the
average~$\varepsilon_b$ approach the asymptotic value
of~$(\mu_b/\mathcal{M})\varepsilon$ given by the first term
of~Eq.~(\ref{eq:e_b}). However, the width of the
$\varepsilon_b$~distribution, determined by the last term of this
equation, increases with rising~$\varepsilon$. The ratio of this width
to the mean value of~$\varepsilon_b$ decreases as~$\varepsilon^{-1/2}$.

\subsection{Spin-dependent scattering}
\label{sec:difmol_sym}

When at least one of the nuclei bound in~$BC$ (e.g., the nucleus~$b$) is
identical with the nucleus~$a$, it is necessary to consider
spin-dependent reactions~(\ref{eq:nuc_scatt_el})
and~(\ref{eq:nuc_scatt_sf}). Let us introduce the following notation
\begin{equation}
  \begin{split}
    \label{eq:spins}
    &\vec{\mathcal{J}} \equiv\vec{s}_{\mu}
    +\vec{s}_a +\vec{s}_b+ \vec{s}_c \,, \qquad  \\
    &\vec{F} \equiv\vec{s}_{\mu}+\vec{s}_a \,, \qquad 
    \vec{I}  \equiv \vec{s}_b +\vec{s}_c \,, \\
    &\vec{S}_b \equiv \vec{F} +\vec{s}_b \,, \qquad 
    \vec{S}_c  \equiv \vec{F} +\vec{s}_c \,, \\
    &\vec{\mathcal{J}} = \vec{F}+\vec{I} 
    = \vec{S}_b + \vec{s}_c = \vec{s}_b+\vec{S}_c \,,
  \end{split}
\end{equation}
where~$\vec{s}_{\mu}$, $\vec{s}_a$, $\vec{s}_b$, and~$\vec{s}_c$, are
the spins of the muon and of the nuclei $a$, $b$, and~$c$, respectively.
Thus, $\vec{\mathcal{J}}$ is the total spin of $a\mu+BC$\@ system
and~$\vec{I}$ is the total nuclear spin of the molecule. It is assumed
that~$\vec{\mathcal{J}}$ is conserved in the scattering since the
spin-orbit interaction is very weak. Also, it is assumed that the
spin~$\vec{S}_b$~($\vec{S}_c$) is conserved in local collisions
of~$a\mu$ with the nucleus~$b$~($c$) bound in~$BC$\@.

If the isotope~$c$ is different from $a$ and~$b$, the molecule~$BC$\@ is
asymmetric and its parity is not definite. Thus, the directions of the
nuclear spins~$\vec{s}_b$ and $\vec{s}_c$ are independent of each other,
and a~unique spin~$\vec{S}_b$ of the subsystem $a\mu+b$ is assigned to
the initial~$\xi_i$ and final~$\xi_f$ spin states of the system
$a\mu+BC$ (with fixed values of~$F$ and~$F'$). These states can be
written down as follows:
\begin{equation}
  \label{eq:spin_asym}
  \begin{split}
    \xi_i(\vec{S}_b,\vec{s}_c;F) & =
    \xi_{a\mu-b}(\vec{S}_b;F)\,\xi_c(\vec{s}_c) \,, \\
    \xi_f(\vec{S}_b,\vec{s}_c;F') & =
    \xi_{a\mu-b}(\vec{S}_b;F')\,\xi_c(\vec{s}_c) \,,
  \end{split}
\end{equation}
where $\xi_{a\mu-b}$ and $\xi_c$ are the eigenfunctions of the conserved
spins~$\vec{S}_b$ and~$\vec{s}_c$, respectively. In this case, the total
wave functions take the form
\begin{equation}
  \label{eq:wave_spin_asym}
  \begin{split}
    &\psi_0 = \phi_i(\vec{r}_{\mu})\, \varPhi_0(\vec{R})\, 
    \exp(i\vec{k}_0\cdot\vec{r})\,\xi_i(\vec{S}_b,\vec{s}_c;F)\,, \\
    &\psi_n = \phi_f(\vec{r}_{\mu})\, \varPhi_n(\vec{R})\, 
     \exp(i\vec{k}_n\cdot\vec{r})\,\xi_f(\vec{S}_b,\vec{s}_c;F')\,, 
  \end{split}
\end{equation}
at large distances~$r$. In order to obtain the correct
molecular-scattering amplitudes, symmetrization of the
functions~$\psi_0$,~$\psi_n$ over the two identical particles~$a$
and~$b$ should be performed. As a~result, we obtain the molecular
amplitude~$\mathcal{F}_{0n}^{(b)}$ which is expressed
by~Eq.~(\ref{eq:amp_mol_expan2}) with~$f_{if}^{(b)}$ replaced by the
spin-dependent nuclear amplitude~$f_{FF'}^{S_b}$ for the
process~(\ref{eq:nuc_scatt_el}) or~(\ref{eq:nuc_scatt_sf}). The
calculated amplitudes~$f_{FF'}^{S_b}$ are already symmetrized over the
identical nuclei~\cite{buba87,brac89a,brac89,brac90}. Similarly, the
total molecular cross section~$d\sigma^{S_b}_{0n}/d\varOmega$ for
$a=b\neq c$ is given by Eq.~(\ref{eq:x_mol_asym}) with~$f_{if}^{(b)}$
replaced by~$f_{FF'}^{S_b}$. In the case of spin-flip reaction, we
substitute $f_{if}^{(c)}=0$ in~Eq.~(\ref{eq:x_mol_asym}), because this
process is very weak when the isotope~$b$ is different
from~$a$~\cite{cohe91c}. The cross
section~$d\sigma^{S_b}_{0n}/d\varOmega$ can be averaged over the
projections of spin~$S_b$, which gives the mean cross
section~$d\sigma_{0n}/d\varOmega$.

When all the hydrogen isotopes are identical, the molecule~$BC$\@ is
symmetric and its initial and final parities~$P_I$ and~$P_I'$ are
definite. Therefore, the values $I$ and~$I'$ of the molecular spin are
definite. The total spin functions of $a\mu+BC$ are now eigenstates
of~$\vec{\mathcal{J}}$ with fixed values of $F$~and~$I$ (or
$F'$~and~$I'$). Thus, the total spin state is determined by four quantum
numbers: the absolute value~$\mathcal{J}$ of the total
spin~$\vec{\mathcal{J}}$, its projection~$\mathcal{J}_z$, $F$, and~$I$.
The initial and final total wave functions of the system with $a=b=c$
are
\begin{equation}
  \label{eq:wave_spin_sym}
  \begin{split}
    &\psi_0 = \phi_i(\vec{r}_{\mu})\, \varPhi_0(\vec{R})\, 
    \exp(i\vec{k}_0\cdot\vec{r})\,\xi_i(\vec{\mathcal{J}};F,I)\,, \\
    &\psi_n = \phi_f(\vec{r}_{\mu})\, \varPhi_n(\vec{R})\, 
     \exp(i\vec{k}_n\cdot\vec{r})\,\xi_f(\vec{\mathcal{J}};F',I')\,, 
  \end{split}
\end{equation}
with the following condition to be satisfied 
\begin{equation}
  \label{eq:spin_F_plus_I}
  \vec{F}+\vec{I} = \vec{F}'+\vec{I}' = \vec{\mathcal{J}} \,.
\end{equation}
The functions~$\xi_i$ and~$\xi_f$ contain contributions from different
states of the operator~$\vec{S}_b$. They can be expanded as follows:
\begin{equation}
  \label{eq:spin_sym}
  \begin{split}
    \xi_i(\vec{\mathcal{J}};F,I) &= 
    \sum_{\vec{S}_b} C_b(\vec{S}_b,\vec{s}_c;F,I)\,
    \xi_{a\mu-b}(\vec{S}_b)\,\xi_c(\vec{s}_c) \,,  \\
    \xi_f(\vec{\mathcal{J}};F',I') &= 
    \sum_{\vec{S}_b} C_b'(\vec{S}_b,\vec{s}_c;F',I')\,
    \xi_{a\mu-b}(\vec{S}_b)\,\xi_c(\vec{s}_c) \,, 
  \end{split}
\end{equation}
with $\vec{s}_c$ subject to the condition:
$\vec{S}_b+\vec{s}_c=\vec{\mathcal{J}}$. The factors
$C_b(\vec{S}_b,\vec{s}_c;F,I)$ and $C_b'(\vec{S}_b,\vec{s}_c;F',I')$ are
obtained by the multiple use of the Clebsh-Gordan coefficients. The
expansions of the total spin functions in terms
of~$\xi_{a\mu-b}(\vec{S}_b)$ are necessary since the presented method is
based on knowledge of the three-body scattering
amplitudes~$f_{FF'}^{S_b}$ evaluated for fixed values of~$S_b$.

After performing a~symmetrization of the total potential
$V^{(b)}+V^{(c)}$ and of the wave functions~(\ref{eq:wave_spin_sym})
over the three identical nuclei and proceeding as in
Sec.~\ref{sec:impulse_approx} for the spinless case, one obtains
\begin{equation}
  \label{eq:amp_3sym_b} 
  \begin{split}
    \mathcal{F}_{0n}^{(b)} = \,
    &\frac{\mathcal{M}}{\mu} \, \sum_{\vec{S}_b} 
    C_b(\vec{S}_b,\vec{s}_c;F,I)\, C_b'(\vec{S}_b,\vec{s}_c;F',I')
    f_{FF'}^{S_b} \\
    &\times \int d^{\,3}R\, \varPhi_n^*(\vec{R}) 
    \exp(i\beta\vec{q}\cdot\vec{R})\, \varPhi_0(\vec{R}) 
  \end{split}  
\end{equation}
In the derivation of Eq.~(\ref{eq:amp_3sym_b}), it has been assumed that
the three nuclei are never close together, i.e., the nucleus~$c$ is only
a~distant spectator when $a\mu$ collides with the nuclei~$b$. As
a~result, the molecular amplitude~(\ref{eq:amp_3sym_b}) is expressed in
terms of the three-body amplitudes~$f_{FF'}^{S_b}$. By
employing~Eq.~(\ref{eq:amp_3sym_b}), the total molecular amplitude takes
the form
\begin{equation}
  \label{eq:amp_3sym} 
  \begin{split}
    \mathcal{F}_{0n} = & 
    \frac{\mathcal{M}}{\mu} \,\mathfrak{F}_{FF'}(K,K') 
    \int d^{\,3}R\, \varPhi_n^*(\vec{R}) \bigl[ 
    \exp(i\beta\vec{q}\cdot\vec{R}) \bigr. \\ 
    &+ \bigl. P_I P_{I'} \exp(-i\beta\vec{q}\cdot\vec{R})\, 
    \varPhi_0(\vec{R})\bigr] ,
  \end{split}  
\end{equation}
where $\mathfrak{F}_{FF'}(I,I')$ is given as
\begin{equation}
  \label{eq:amp_f_ff'_ii'} 
  \mathfrak{F}_{FF'}(I,I') \equiv \sum_{\vec{S}_b} 
  C_b(\vec{S}_b,\vec{s}_c;F,I)\, C_b'(\vec{S}_b,\vec{s}_c;F',I')
    \, f_{FF'}^{S_b} \,.
\end{equation}
This result is independent of a~choice of nucleus~$b$ because of the
symmetry: $b\leftrightarrow c$, $\vec{S}_b\leftrightarrow \vec{S}_c$,
and $\vec{s}_c\leftrightarrow \vec{s}_b$. Using the expansion of the
plane wave in terms of the spherical harmonics
in~Eq.~(\ref{eq:amp_3sym}) and taking into account that %
$\left[1+P_I P_{I'}(-1)^l\,\right]=2$ for every allowed rotational
transition, we obtain
\begin{equation}
  \label{eq:x_mol_sym2}
  \frac{d\sigma_{0n}}{d\varOmega} = 
  4 \left(\dfrac{\mathcal{M}}{\mu}\right)^{\!2  }\,
  \overline{\bigl| \mathfrak{F}_{FF'}(K,K')\bigr|^2} \;
  \frac{k_n}{k_0} \sum_l \mathcal{W}_{K'lK} \, 
  \mathcal{D}_{\nu l}^2(\beta q) ,
\end{equation}
in the case $a=b=c$. Since, in experiments, both the muonic atoms and
the target molecules are not polarized, the cross section is averaged
over~$\mathcal{J}$ and~$I$ and summed over~$I'$ (for fixed $K$ and
$K'$), which is denoted by the horizontal bar over the squared
amplitude~$|\mathfrak{F}_{FF'}|^2$.

\subsection{Electron screening corrections to molecular cross sections}
\label{sec:screen_corr}

The differential molecular cross sections derived in the previous
section include only the muonic-atom interaction with nuclei. It is
necessary, however, to include electron screening effects in $a\mu$
scattering from molecules. At $\varepsilon\lesssim{}1$~eV, the relative
velocity of $a\mu$ and~$BC$\@ is smaller by several orders of magnitude
than the muon velocity in $a\mu$ and is also smaller than the electron
velocity in the molecule. Therefore, it is possible to introduce an
effective electron-screening potential, which is obtained by averaging
the Coulomb interaction between $a\mu$ and the electrons over the muon
and the electron coordinates. The range of a~dominant fraction of the
$a\mu$\nobreakdash-$b$ potential is smaller than about
20~$a_{\mu}$~\cite{vini82}. On the other hand, the $a\mu$ interaction
with the electrons is important at distances on the order of the Bohr
radius $a_e\approx$~207~$a_{\mu}$ of the electronic hydrogen atom. Thus,
$a\mu$~collision with an ordinary molecule can be described as
scattering on the two potentials with very different ranges.

The effective screening potential~$V_\text{el}$ for $a\mu$ scattering
from hydrogenic molecules has the following form~\cite{adam86b,adam89}:
\begin{equation}
  \label{eq:screen_pot}
  \begin{split}
    V_\text{el} = & 
    -\frac{\mathcal{C}\,\eta^3}{a_e^3\,(1+S_\eta^2)}
    \Biggl\{
    \exp\left(-\frac{2\eta}{a_e}\bigl|\vec{r}+
      \beta_b\vec{R}\bigr|\right) \\
    & +2 S_\eta
    \exp\left[-\frac{\eta}{a_e}
      \Bigl(\bigl|\vec{r}+\beta_b\vec{R}\bigr| + 
      \bigl|\vec{r}-\beta_c\vec{R}\bigr|\Bigr)\right] \\
    & +\exp\left(-\frac{2\eta}{a_e}\bigl|\vec{r}-
      \beta_c\vec{R}\bigr|\right)
    \Biggr\} ,
  \end{split}
\end{equation}
in which 
\begin{equation*}
  \mathcal{C} = 2\,\varkappa_{\mu} + 8.4\,\sqrt{\,m_e} \,, 
  \quad
  \varkappa_{\mu} = (M_a - m_\mu)/(M_a + m_\mu) \,,
\end{equation*}
and
\begin{equation*} 
  S_\eta = 
  \bigl(1 + w_{\eta} + \tfrac{1}{3}\,w_{\eta}^2\bigr)
  \exp(-w_{\eta}) , \quad   w_{\eta} = \frac{\eta R_0}{a_e} , 
  \quad \eta=1.2
\end{equation*}
The electronic correction to the process~(\ref{eq:mol_scatt_el}) is
calculated using the first Born approximation. The total molecular
amplitude~$\mathcal{F}_{0n}^{\text{~mol}}$ is now equal to sum of the
nuclear amplitudes and the screening
amplitude~$\mathcal{F}_{0n}^{\text{el}}$
\begin{equation}
  \label{eq:amp_mol-el}
  \mathcal{F}_{0n}^{\text{~mol}}(\vec{k}_0,\vec{q}) =
  \mathcal{F}_{0n}^{(b)}(\vec{k}_0,\vec{q}) +
  \mathcal{F}_{0n}^{(c)}(\vec{k}_0,\vec{q}) +
  \mathcal{F}_{0n}^{\text{el}}(\vec{q})  \,,    
\end{equation}
where 
\begin{equation}
  \label{def:amp-el}
  \begin{split}
    \mathcal{F}_{0n}^{\text{el}}(\vec{q}) = -\frac{\mathcal M}{2\pi}
    \int & d^3r\, d^3R \, \exp(-i\vec{q}\cdot\vec{r})\, \\
    &\times \varPhi_n^*(\vec{R})\,V_\text{el}(\vec{r},\vec{R})\, 
    \varPhi_0(\vec{R}) \,. 
  \end{split}   
\end{equation}
The calculated amplitude~$\mathcal{F}_{0n}^{\text{el}}$ falls rapidly
when $qa_e\gtrsim{}1$, which occurs even for the lowest rotational
excitations of a~hydrogenic molecule. Therefore, it is sufficient to
take this amplitude into account only for the strictly elastic
scattering.  The isotopic exchange~(\ref{eq:nuc_scatt_ex}) and the
strong spin-flip~(\ref{eq:nuc_scatt_sf}) reactions are due to the
exchange of the muon between two nuclei taking part in direct collision.
Therefore, $a\mu$~scattering from electrons cannot cause these
reactions.The first non-vanishing screening corrections to the spin-flip
or isotopic-exchange cross sections appear only in the distorted wave
Born approximation.

A~further evaluation of the screening corrections to the scattering
amplitudes should be performed numerically. With regard to the elastic
processes, these corrections are very significant. For example,
at~$\varepsilon\to{}0$, the screening
amplitude~$\mathcal{F}_{0n}^\text{el}$ for $p\mu+\text{H}_2$ elastic
scattering is comparable to the corresponding $p\mu+p$ scattering
amplitude. In this limit, the relative screening corrections to the
spin-flip or isotopic-exchange amplitudes are on the order of~10\%. At
$\varepsilon\gtrsim{}1$~eV, screening effects practically vanish for all
processes.

\section{Examples of molecular cross sections}
\label{sec:example_mol}

In this section some typical examples of the molecular cross sections
are shown. The nuclear scattering amplitudes given in
Refs.~\cite{buba87,brac89a,brac89,brac90,adam92,chic92} are used as the
input for computation of the molecular differential cross sections.
These amplitudes are first averaged over the internal motion of nuclei
inside the target molecules, according to~Eq.~(\ref{eq:amp_avg_J}).
\begin{figure}[htb]
  \begin{center}
    \includegraphics[width=7cm]{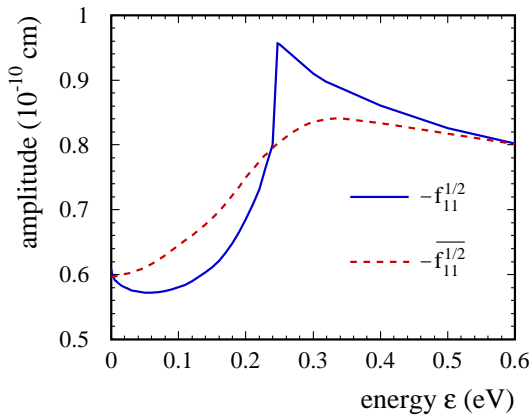}
    \caption{(Color online) Amplitude~$f_{11}^{1/2}$ 
      and average amplitude~$\overline{f_{11}^{1/2}}$ 
      for $p\mu(F=0)+p$ scattering versus collision 
      energy~$\varepsilon$ in the molecular~c.m.s..
      \label{fig:avgam_p11}}
  \end{center}
\end{figure}
\begin{figure}[htb]
  \begin{center}
    \includegraphics[width=7cm]{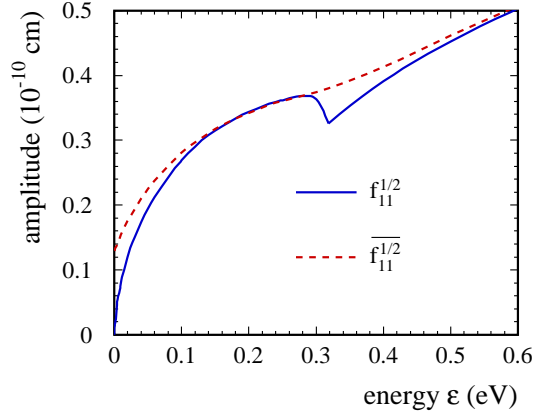}
    \caption{(Color online) The same as in Fig.~\ref{fig:avgam_p11}, 
      for $t\mu(F=0)+t$ scattering.
      \label{fig:avgam_t11}}
  \end{center}
\end{figure}
In~Figs.~\ref{fig:avgam_p11} and~\ref{fig:avgam_t11}, the averaged
nuclear amplitudes~$\overline{f_{11}^{1/2}}$ for the elastic scattering
$p\mu(F=0)+$H$_2$ and $t\mu(F=0)+$T$_2$ are shown. The input nuclear
amplitudes~$f_{11}^{1/2}$ are plotted versus collision
energy~$\varepsilon$ in the molecular c.m.s., using the high-energy
asymptotic relation $\varepsilon=(\mathcal{M}/\mu_b)\,\varepsilon_b$.
The cusp in the elastic cross section $p\mu+p$, located at the spin-flip
threshold, is smeared out after the averaging over proton motion
in~H$_2$. Although the vibrational quantum for~T$_2$ is smaller than
that for~H$_2$, smoothing of the amplitude for $t\mu+t$ elastic
scattering is also important, owing to strong changes of its value
within the energy interval of~0.1~eV. This is particularly visible in
the cusp region and at $\varepsilon\to{}0$. This smearing strongly
affects the molecular cross sections, which are expressed by the squared
amplitudes. The elastic cross sections for $d\mu+d$ scattering are quite
flat at the lowest energies~\cite{buba87,brac89,brac90}. As a~result,
differences between the amplitudes~$f_{if}^S$ and~$\overline{f_{if}^S}$
are much smaller than in the protium or tritium case. The role of
smearing effects were investigated during final a~analysis of the PSI
diffusion data~\cite{abbo97}. A~spectacular improvement of the fits to
the data, especially for $p\mu$ diffusion in~H$_2$, was achieved when
the averaged nuclear amplitudes~$\overline{f_{if}^S}$ were used for the
calculations of the molecular cross sections. This mainly concerns the
elastic cross sections, as smoothing effects are generally much smaller
in the case of spin-flip or isotopic-exchange amplitudes, which weakly
depend on the energy
below~1~eV~\cite{buba87,brac89a,brac89,adam92,chic92}.

Electron-screening and molecular-binding effects are clearly seen in the
molecular differential cross sections. The range of the screening
potential~(\ref{eq:screen_pot}) is on the order of~$a_e$, so that the
condition $k_0{}a_e\sim{}1$ is fulfilled already at
$\varepsilon\sim$~0.001~eV. Many partial waves begin to contribute
significantly to the screening amplitude~(\ref{def:amp-el})
above~0.001~eV. As a~result, the molecular cross sections are
anisotropic even at very low energies. Moreover, scattering from
a~molecule is connected with different rotational transitions, which
additionally leads to a~complicated angular distribution of the
scattered atoms. This is in contrast to $a\mu$~scattering from a~bare
hydrogen nucleus, where few partial waves contribute significantly to
the nuclear cross sections below~100~eV. In all of the three-body cases,
the $s$-wave cross section describes scattering below 0.1--1~eV well,
which is therefore isotropic in the nuclear~c.m.s..
\begin{figure}[htb]
  \begin{center}
    \includegraphics[width=7cm]{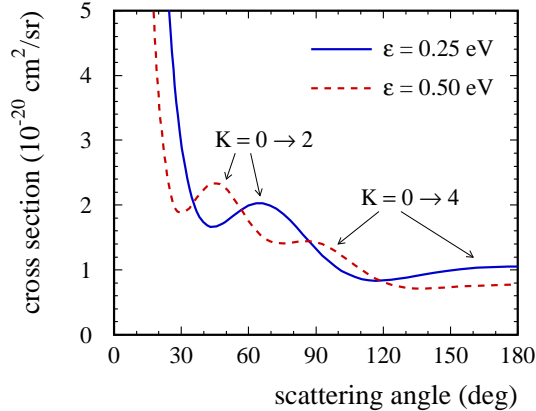}
    \caption{(Color online) Differential cross section 
      for $p\mu(F=0)$ scattering from a~H$_2(K=0)$ molecule 
      versus scattering angle~$\vartheta$.
      \label{fig:xang_ppp}}
  \end{center}
\end{figure}
The cross section $d\sigma_{0n}/d\varOmega$ for $p\mu$ scattering from
a~ground-state H$_2$ molecule is presented in~Fig.~\ref{fig:xang_ppp},
for $\varepsilon=0.25$~eV and $\varepsilon=0.5$~eV. Scattering at the
angles $\vartheta\lesssim{}30^\circ$ is dominated by $p\mu$ scattering
from the electron cloud. The peaks at greater angles are due to the
rotational transitions $K=0\to{}K'=2$ and $K=0\to{}K'=4$. The angular
positions of the scattering peaks change with the variation of $p\mu$
energy.

\begin{figure}[htb]
  \begin{center}
    \includegraphics[width=6.5cm]{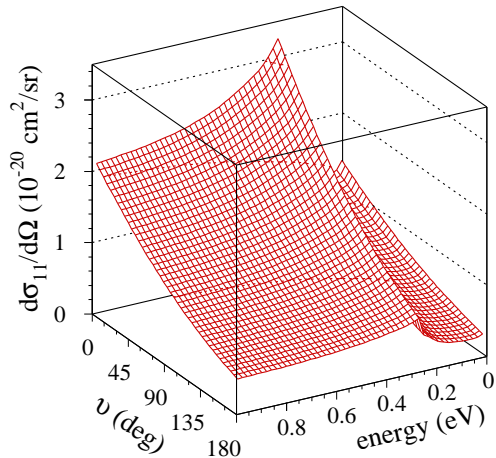}
    \caption{(Color online) Doubled differential cross section for
      $p\mu(F=0)+p$ scattering versus~$\varepsilon$ and~$\vartheta$.
      \label{fig:nucdif_ppp11}}
  \end{center}
\end{figure}

\begin{figure}[htb]
  \begin{center}
    \includegraphics[width=6.5cm]{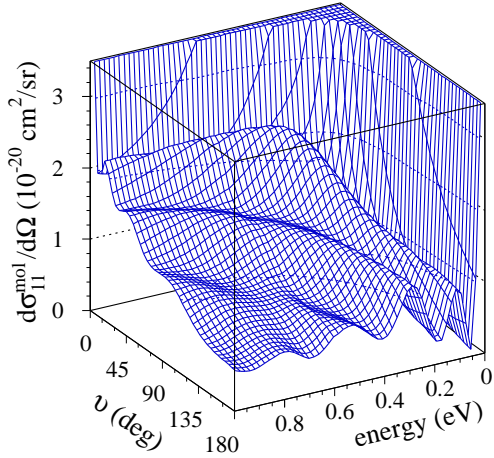}
    \caption{(Color online) Differential cross section for
      $p\mu(F=0)$+H$_2(K=0)$ scattering versus~$\varepsilon$ 
      and~$\vartheta$.
      \label{fig:xdif_ppp11}}
  \end{center}
\end{figure}

The differential cross sections for $p\mu$ scattering from a~free proton
and from a~H$_2$~molecule are plotted in~Figs.~\ref{fig:nucdif_ppp11}
and~\ref{fig:xdif_ppp11} as functions of the collision
energy~$\varepsilon$ and the scattering angle~$\vartheta$ in the
molecular~c.m.s..
Figure~\ref{fig:nucdif_ppp11} illustrates the cross section for the
elastic $p\mu+p$ scattering, multiplied by~2. The mass of the target
particle is, however, set to the H$_2$~mass, for the sake of comparison
with this molecular target. Only the $s$~wave contributes to the
$p\mu+p$ cross section in the considered energy interval. However, the
scattering is not isotropic in the molecular~c.m.s.. A~cusp is apparent
at the energy of the spin-flip threshold.
The surface in~Fig.~\ref{fig:xdif_ppp11} describes the corresponding
cross section $d\sigma_{0n}/d\varOmega$ (cut above a~value of
$3.5\times{}10^{-20}$~cm$^2$/sr) for the ground-state H$_2$~molecule.
The electron-screening contribution to the differential cross section is
clearly seen at all angles for $\varepsilon\lesssim{}0.05$~eV and only
as a~forward peak at higher energies. The lowest rotational transitions
can be distinguished in this plot. Also a~smearing of the molecular
cross section due to the proton internal motion in~H$_2$ is visible,
especially in the spin-flip threshold region. The sections of the
surfaces shown in Figs.~\ref{fig:nucdif_ppp11} and~\ref{fig:xdif_ppp11}
with the plane~$\varepsilon=1$~eV are already quite similar, apart from
the forward scattering. Thus, the molecular scattering at larger
energies and angles, which contains contributions from many
rotational-vibrational transitions, approaches the ``doubled'' nuclear
scattering.
\begin{figure}[htb]
  \begin{center}
    \includegraphics[width=7cm]{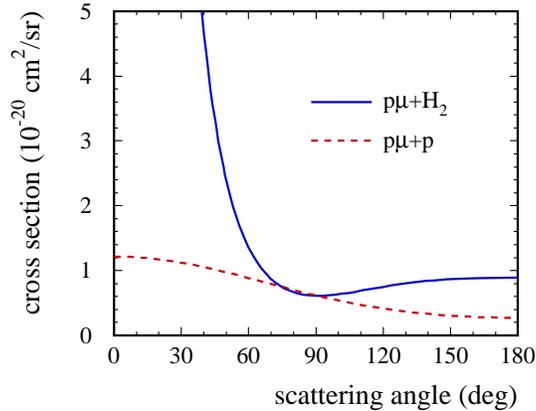}
    \caption{(Color online) Angular dependence of the differential 
      cross sections for $p\mu(F=0)$ scattering on a~proton (doubled) 
      and on a~H$_2(K=0)$, at~$\varepsilon=0.1$~eV.
      \label{fig:xangcmp_ppp0.1}}
  \end{center}
\end{figure}
\begin{figure}[htb]
  \begin{center}
    \includegraphics[width=7cm]{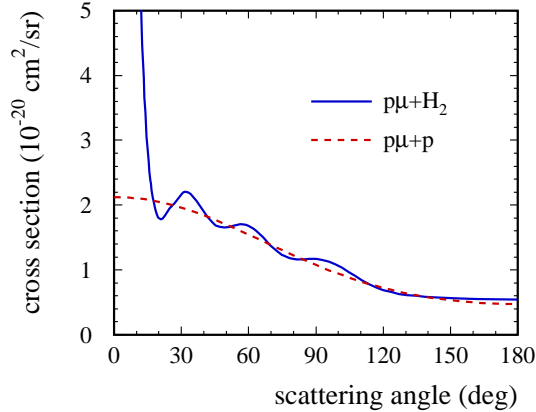}
    \caption{(Color online) The same as
      in~Fig.~\ref{fig:xangcmp_ppp0.1}, at~$\varepsilon=1$~eV.
      \label{fig:xangcmp_ppp1.0}}    
  \end{center}
\end{figure}
For a~better comparison, the angular dependence of the doubled nuclear
and the molecular cross sections are shown
in~Figs.~\ref{fig:xangcmp_ppp0.1} and~\ref{fig:xangcmp_ppp1.0}, at
a~fixed collision energy. For~$\varepsilon=0.1$~eV, the two cross
sections are very different at all angles. For $\varepsilon=1$~eV, these
cross sections are already quite similar at larger angles
$\vartheta\gtrsim{}20^\circ$. One sees that there are only small
rotational oscillations of the molecular curve around the doubled
nuclear curve.

\begin{figure}[htb]
  \begin{center}
    \includegraphics[width=6.5cm]{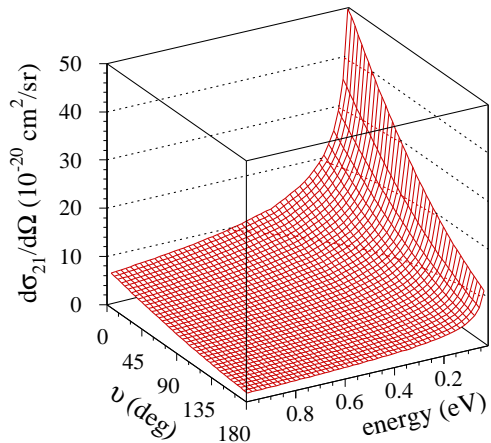}
    \caption{(Color online) Doubled differential cross section for 
      the downwards spin-flip reaction in $p\mu(F=1)$ scattering from 
      a~proton versus~$\varepsilon$ and~$\vartheta$.
      \label{fig:nucdif_ppp21}}
  \end{center}
\end{figure}
\begin{figure}[htb]
  \begin{center}
    \includegraphics[width=6.5cm]{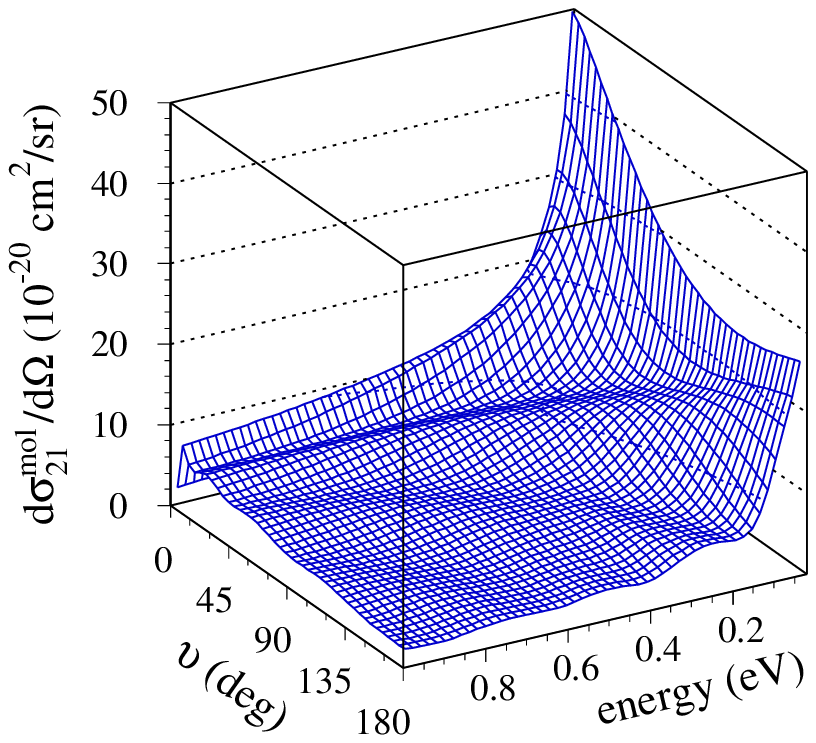}
    \caption{(Color online) Differential cross section for the 
      downwards spin-flip reaction in $p\mu(F=1)$ scattering from 
      a~H$_2(K=0)$ versus~$\varepsilon$ and~$\vartheta$.
      \label{fig:xdif_ppp21}}
  \end{center}
\end{figure}
The differential cross sections for the downwards spin-flip reaction in
$p\mu(F=1)$ scattering from a~proton and from a~H$_2$~molecule are shown
in~Figs.~\ref{fig:nucdif_ppp21} and~\ref{fig:xdif_ppp21}.
Electron-screening and molecular-binding effects in the spin-flip
reactions are not so important as in the case of spin-conserving
scattering (cf.~Figs.~\ref{fig:xdif_ppp11} and~\ref{fig:xdif_ppp21}),
owing to higher momentum transfers. Larger differences between the
nuclear and the molecular spin-flip cross sections appear mainly at
small collision energies and small angles.

Simulations of experiments performed in gaseous hydrogenic targets
require knowledge of the differential cross sections for the molecular
processes~(\ref{eq:mol_scatt}). These molecular cross sections have been
computed and stored as computer files. They have been applied for
planning and interpreting many experiments in H-D gaseous targets. For
example, optimal conditions for studies of $\mu^{-}$ nuclear capture in
$p\mu$~\cite{kamm00,kamm01} and for the measurement of the Lamb shift in
$p\mu$ atoms~\cite{kott01,pohl01a,pohl05} created in H$_2$ targets have
been determined using the calculated molecular cross sections. These
experiments are now underway at the Paul Scherrer Institute.

\section{Conclusions}

A~method of calculating the differential cross sections for low-energy
muonic atom scattering from hydrogenic molecules has been developed.
This method directly uses the corresponding amplitudes for muonic atom
scattering from hydrogen-isotope nuclei, calculated within the framework
of the adiabatic method for the three-body problem with the Coulomb
interaction. Thus, the presented method naturally includes the angular
and energy dependence of the three-body amplitudes in the scattering
from hydrogenic molecules. Since, in many cases, the considered
three-body scattering amplitudes depend strongly on the collision energy
within the interval $\lesssim$~0.1-eV, a~broad distribution of the
nucleus kinetic energy in a~hydrogenic molecule is taken into account.
The molecular vibrations are described in the harmonic approximation.
Therefore, the evaluated cross sections are valid below a~few~eV.

For a~fixed energy~$\varepsilon$ of a~muonic atom collision with
a~hydrogenic molecule, the calculated collision energy~$\varepsilon_b$
in the system consisting of the atom and a~single hydrogen-isotope
nucleus has a~wide spectrum. At~$\varepsilon\to{}0$, this spectrum
reveals a~shape of the kinetic energy distribution of the nucleus in
a~given rotational-vibrational state of the molecule. This effect is
very significant, even for the molecular ground state, as the energy of
zero-point vibration in hydrogenic molecules is quite large. The width
of the $\varepsilon_b$~spectrum for the lightest H$_2$ molecule is on
the order of~0.1~eV. At higher~$\varepsilon$, this width is even larger.
As a~result, the three-body amplitudes are strongly smoothed when
calculating the molecular cross sections. This effect and the energy and
angular dependence of the three-body amplitudes are included in the
calculated set of the differential cross sections for low-energy
scattering of 1$S$~muonic hydrogen atoms from hydrogenic molecules.
These are the only to date theoretical cross sections which give good
agreement with many experiments involving $p\mu$ and $d\mu$ scattering
in gaseous H-D targets. The presented method can also be applied for
scattering of other ground-state exotic atoms or neutrons from
hydrogenic targets.


\bibliography{adamczak}

\end{document}